# Elimination of Central Artefacts of L-SPECT with Modular Partial Ring Detectors by Shifting Center of Scanning

Manu Francis, Murat Tahtali and Mark R Pickering

*Abstract*—The Lightfield Single Photon Emission Computed Tomography (L-SPECT) system is developed to overcome some of the drawbacks in conventional SPECT by applying the idea of plenoptic imaging. This system displayed improved performance in terms of reduced information loss and scanning time when compared to the SPECT system which has a conventional collimator. The SPECT system is transformed into L-SPECT system by replacing the conventional collimators with micro-range multi-pinhole arrays. The field of view (FOV) of the L-SPECT system is enhanced by reshaping the detector head into ring-type by tiling small detector modules. The L-SPECT system with modular partial ring detectors (MPRD L-SPECT) exhibits cylindrical artefacts during volumetric reconstruction. Hence, here the work is focused to remove the cylindrical artefacts in the reconstruction of the examined objects by changing the scanning orbit. The enhancement is done such that the center of scanning of the L-SPECT system with MPRD L-SPECT is shifted at different values. The reconstruction quality of MPRD L-SPECT with and without center shifting is evaluated in terms of the Full Width at Half Maximum (FWHM) and Modulation Transfer Function (MTF). Moreover, the visual comparison is also examined. The results indicate that center shifting of MPRD L-SPECT overcomes the problem of the central artefact with improved FWHM. By increasing MPRD's scanning center shift gap, the spatial resolution can be further improved.

## I. INTRODUCTION

THE Single Photon Emission Computed Tomography is a nuclear imaging technology to analyze the functionality of human and animal organs [1]. The conventional SPECT systems were grouped based on the type of collimator used in these systems. The main types of collimators employed in SPECT systems are parallel hole, converging, diverging and pinhole collimators. These common SPECT systems experience the shortcomings of high resolution and low sensitivity [2]. The reduced field of view (FOV) and the usage of collimators are main causes of low sensitivity issue [3]. Therefore, large number of projection images at different view angles are required for proper reconstruction. Moreover, higher scanning time for each view angles is required. However, this increased scanning time leads to the introduction of motion artefacts in the reconstructed objects. The primary motivation behind the research in SPECT imaging area is to increase the sensitivity as well as the resolution by enhancing the collimator structure and reconstruction algorithms [4]-[12]. The Lightfield SPECT (L-SPECT) system is designed to reduce the above-mentioned problems. The L-SPECT is a revamped version of SPECT system designed by employing plenoptic imaging concept [13]. The plenoptic imaging is a technique to capture images along with the direction. Therefore, in L-SPECT, the system design makes the system capable to identify incident gamma ray direction. The conventional collimator system in L-SPECT system is replaced with micro range multi-pinhole arrays. This collimator is placed in-front of the detector to avoid overlapping the projections behind each pinhole. Moreover, each projection represents different view of the scanned phantom at different view angles. Apart from this, the multi-pinhole configuration enhances the overall sensitivity of the L-SPECT system by improving FOV.

The main hardship in L-SPECT imaging is the back-tracing of the incident ray for the reconstruction because incoming ray direction is not fully known. However, the arrangement of pinhole plate helps to locate the corresponding pinhole through which each detected gamma ray passed. This setup reduces complexities to recognize radiation direction for reconstruction. The scanned phantom is volumetrically based on raytracing. The rays are back-traced by launching a ray from detector pixel through corresponding pinhole center towards cube of interest. The Siddon's ray tracing algorithm identifies back-projected ray intersection with voxels. The intersected voxel values updated with the product of ray intersection length with detector pixel value.

The Modular Partial Ring Detector based L-SPECT (MPRD L-SPECT) is introduced to enhance the FOV [15]. In this modified L-SPECT system, modular detectors are arranged in a partial curve structure. This arrangement supports the convergence of the pinholes on the object to be scanned. Moreover, this arrangement helps to cover a large portion of the phantom in each scan and reduces the overall scanning time. However, from the experiments, some reconstruction artefacts are discernable at the center. The main reason behind this is thought to be a sort of Moiré interference due to repeating symmetrical pinhole pattern.

The work in this paper aims to remove these central artefacts in the MPRD L-SPECT by modifying the center of scanning.

M. Francis is with the School of Engineering and IT, University of New South Wales, Canberra, Australia (corresponding author, e-mail: manu.francis@student.adfa.edu.au).

M. Tahtali., is with the School of Engineering and IT, University of New South Wales, Canberra, Australia

M. R. Pickering., is with the School of Engineering and IT, University of New South Wales, Canberra, Australia

During the scanning process, the scanning center of the detector shifted for every step angle. The shift value varies for different pinhole configurations.

Section II describes the experimental tools, for simulation and reconstruction of the proposed work. Comparison results, both qualitative and quantitative, are discussed in section III. Section IV concludes this paper.

## II. MATERIAL AND METHODS

The simulation of the modular partial ring detector-based L-SPECT system with center shifting was implemented in MATLAB$^{TM}$. Different pinhole and detector configurations were carried out through simulations. Results of the MPRD L-SPECT reconstructions with and without center shifting were analyzed both visually and numerically. Full Width at Half Maximum (FWHM) of Gaussian fitted point source response and the modulation transfer function are used for numerical comparison.

### A. Objects used for the study

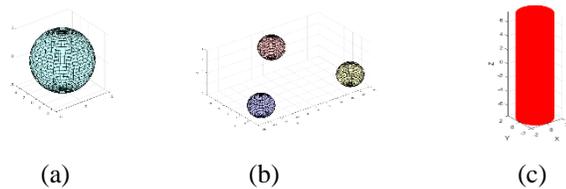

(a) (b) (c)
Fig. 1 Phantoms used for study (a) Sphere (b) 3 Spheres (c) Cylinder

Fig. 1 illustrates the geometric shapes including spheres, cylinders, that were used as radiation sources. The radiation points inside the geometric shapes were generated randomly based on a Monte Carlo simulation. A small sphere doubling as a point source is used for the identification of the point spread function (PSF). The radius of the point source should be less than 10 times of the expected FWHM according to [16], so we selected a radius of 0.2mm [16].

### B. Scanning Simulation

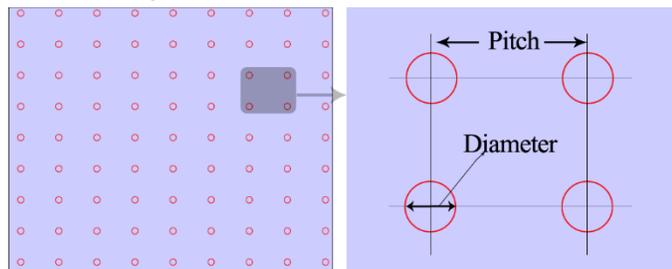

Fig 2. Multipinhole array

The MPRD-LSPECT is simulated by the tiling pinhole plates and detectors. The micro range multi-pinhole configuration is described in Fig. 2. Each tile's width is selected as one pinhole pitch. These tile modules arrange in the shape of a partial circular form as shown in Fig. 3. The width of one pinhole module, and the perpendicular gap between the object center and pinhole module directly contribute to the field of view of the modular partial ring detector.

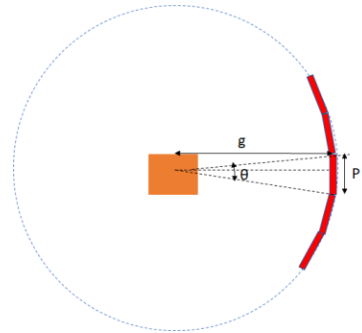

Fig 3. Modular Partial Ring Detector

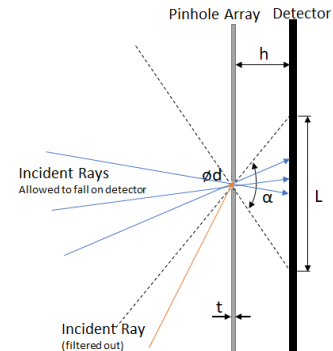

Fig. 4 FOV of pinhole array

Then the center of the detector is shifted based on a fixed value as in Fig. 6(b). The total view angle covered by the modular partial detector $\phi$ is calculated as:

$$\phi = 2N tan^{-1}\left(\frac{P}{2g}\right) \quad (1)$$

$N$ is the total number of detector modules; $P$ is the width of one module and $g$ is the gap between the object center to the multi-pinhole module. Number of scanning $n$ required for complete scanning is:

$$n = \frac{360}{\phi} \quad (2)$$

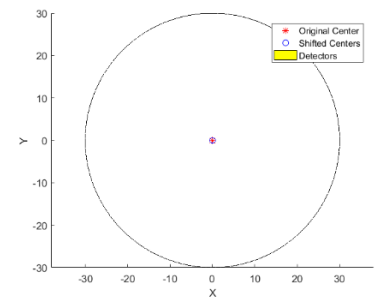

(a)

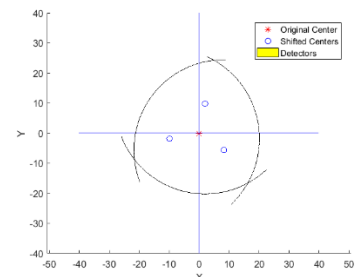

(b)

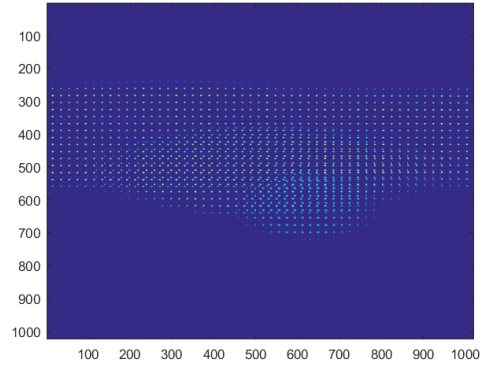

(a)

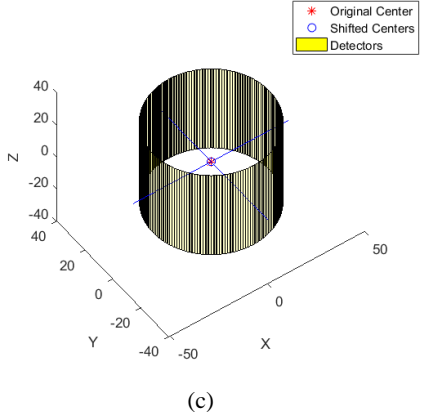

(c)

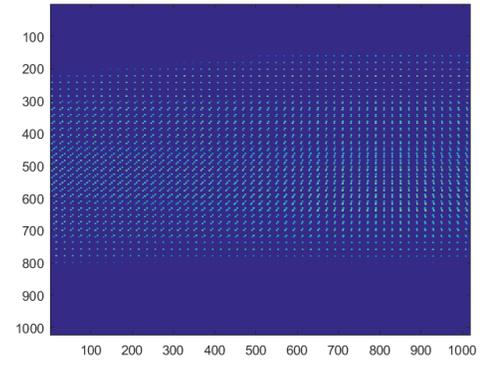

(b)

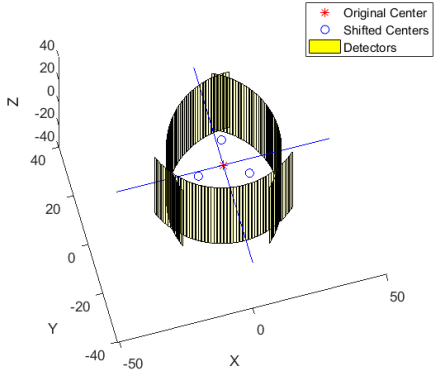

(d)

Fig. 5 Top view of a) MPRD L-SPECT with the center not shifted and b) MPRD L-SPECT with the center shifted, and 3D views of c) MPRD L-SPECT with the center no shifted, and d) MPRD L-SPECT with the center shifted simulation setups.

The center of this modular partial ring detector (MPRD) is then shifted to remove the central artefacts as in Fig. 5. Different configurations of MPRD with varied values of central shifts, pinhole radii and pinhole pitches are selected for the study. The selected pinhole pitch determines the total number of pinholes in the modular detector. The space between the pinhole plate and the detector is adjusted, such that the projection of two adjacent pinholes would not overlap. The gap is calculated as:

$$h = \frac{L.t}{2d} \quad (3)$$

$L$, $t$ and $d$ are the pinhole pitch, the pinhole plate thickness and the pinhole diameter, respectively. Moreover, the field of view is controlled by parameters like diameter and pitch of the pinhole. The field of view (FOV) marked in Fig. 4, of the multi-pinhole array configuration is calculated as [13]:

$$\alpha = 2tan^{-1}\frac{d}{t} \quad (4)$$

The L-SPECT system is simulated by applying a modified Monte Carlo method, so as to introduce randomness in the radiation generation. The radiation points are triggered randomly inside the phantom to be scanned. From these radiation points, rays are sent towards the detector through random locations inside the pinhole's aperture plane [14]. The simulation time is reduced by discarding the radiation generated outside the field of view corresponding to each scan angle.

Fig. 6 Projection images of the phantom comprising of 3 spherical objects scanned with a) MPRD L-SPECT with center shifting of 10 mm, and b) MPRD L-SPECT with no center shifting at projection angle zero.

Moreover, the rays through the pinholes are filtered using the field of view α of the pinhole, to include attenuation through pinhole walls. The intersection of the remaining rays with the detector plane is calculated and identified as pixel locations of the projection image. For every ray intersection at specific pixel of the detector, the value of the corresponding location in the projection image is incremented by one. The projections for all modules in the modular partial ring detector are simulated by rotating one module by the corresponding step angle an:

$$\theta = 2tan^{-1}\frac{P}{2g} \quad (5)$$

Thus, the projection behind one modular partial ring detector is simulated by rotating one module N number of times with step angle θ. *N* is the total number of detector pinhole modules in the modular detector. After completing the projection simulation for one view angle, the modular detector is shifted back and rotated for the next scan angle. Then, the center of the modular detector is shifted again, and the scanning repeated. Fig. 6(a) shows projections of the MPRD L-SPECT with the scanning center shifted.

*C. The reconstruction method*

The scanned object is reconstructed by back-projecting rays from the projections towards the cube of interest. The detector pixels behind the FOV of each pinhole is grouped as a grid. A ray is traced from each non-zero pixel center through the pinhole center of the corresponding grid towards the cube of interest [14]. Siddon's ray tracing algorithm can be used to trace the voxels through which the ray passes [17]. This reconstruction process is repeated for all scan angles and the sum of the reconstructions is calculated to obtain the final reconstructed image.

*D. System configurations used for the study*

Different configurations of central shifted MPRD L-SPECT were used for the analysis. The multi-pinhole arrays for the imaging system were implemented using modules of width equal to 1, 2 and 3 times the value of the pinhole pitch. The simulated planar L-SPECT system has a detector area of 49.152X49.152 mm2 to match the X-ray sensor that we have, thus used as reference. For the purpose of this study, different pinhole pitches having values 0.96mm, 1.92mm, and 2.88mm, are taken as 20, 40 and 60 times the values of sensor pitch, respectively, and the pinhole radius of 0.096mm is twice the sensor pitch value. The space between the pinhole module and the object was chosen as 15mm, 20mm, and 25mm and the phantom is reconstructed into a cube having voxels of value 128x128x128.

## III. RESULTS AND ANALYSIS

*A. Tools used for the analysis*

The reconstruction results of the L-SPECT system having center shifted modular partial ring detector are compared both visually and numerically with the same imaging system having MPRD without shift and with planar MPRD detector. The spatial resolution and the modulation transfer function (MTF) are used for numerical comparison. After reconstruction, central slices of the reconstructed volume were selected for visual inspection and identification of the phantom shapes. The phantom was extracted from the reconstructed volume by thresholding the same with values equal to half of the maximum count in the cube of interest [14].

1) *Point Spread Function:* This analysis is conducted to identify the impulse response of the center shifted MPRD L-SPECT system. The output of an imaging system is the convolution of the input with the PSF of that optical system [18], expressed as:
$$g(x,y) = f(x,y) * h(x,y) \quad (6)$$
where $h(x, y)$ is the PSF. The plot of the counts in the reconstructed image of a point source against the location or voxel will give the point source response. The central slice of the reconstructed volume was used for plotting the count profile. Further, this profile was fitted to a gaussian curve, which is an approximation of the PSF [19].

2) *Spatial Resolution:* The ability of an imaging system to resolve the smallest details in the imaged or scanned object is defined as spatial resolution. Generally, the spatial resolution is calculated as the FWHM of the point source response. The FWHM will be typically 1.4 to 2 times the spatial resolution. FWHM will be calculated from the PSF [16].

3) *Modulation Transfer Function (MTF):* In some specific cases, the same imaging systems can have the same FWHM while scanning different objects. Analysis based on the MTF helps to overcome this issue. The MTF can be used as an analysis tool to quantify resolution and contrast together [20]. The MTF is the frequency domain representation of the PSF and is calculated by taking the Fourier transform of the PSF and extracting its magnitude as:
$$MTF = |FT(PSF)| \quad (7)$$
At frequencies where the MTF values are closer to unity imply better reconstruction at those frequencies.

*B. Reconstruction*

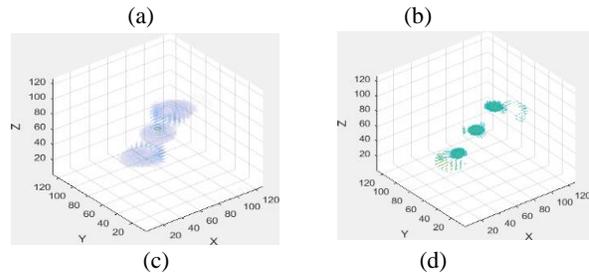

Fig. 7 XZ reconstruction slice for the same phantom using a) MPRD L-SPECT with center shifting of 10 mm and b) MPRD L-SPECT with no center shifting, and the 3D reconstructed volumes for c) MPRD L-SPECT with center shifting of 10 mm and d) MPRD L-SPECT with no center shifting at projection angle zero.

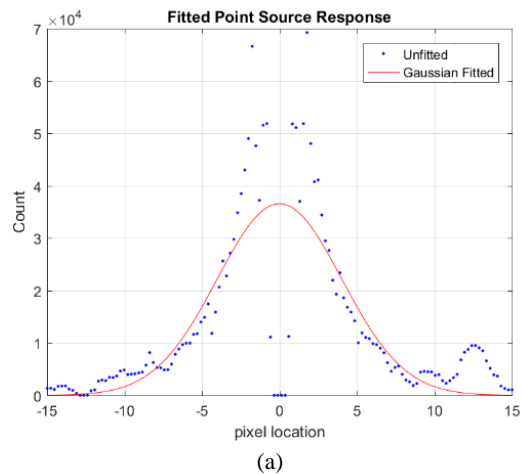

(a)

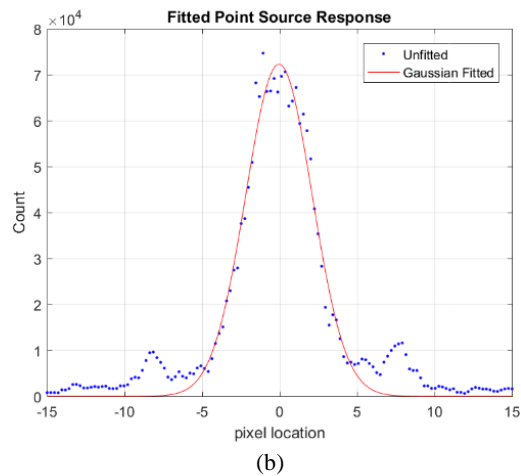

(b)

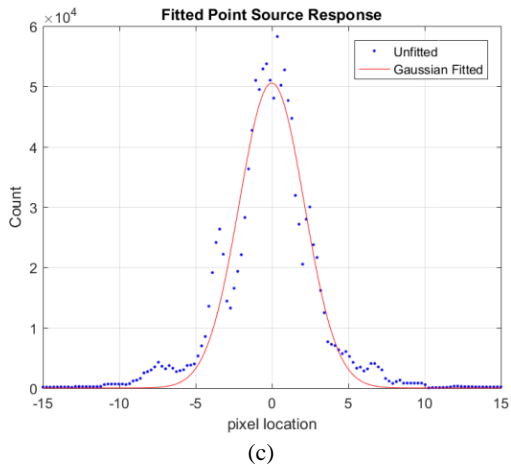

(c)

Fig. 8 Gaussian fitted point source response of a) MPRD L-SPECT with no center shifting b) MPRD L-SPECT with center shifting of 5 mm c) MPRD L-SPECT with center shifting of 10 mm.

The scanned object was reconstructed from projections using the method described in Section II-C. The reconstruction procedure was executed without any preprocessing of projection. The reconstruction results of the phantom with three spherical objects are shown in Fig. 7. The values of the pinhole pitch, radius and module width of the modular partial ring detectors control the quality of the reconstruction.

TABLE I
FWHM comparison of different MPRD L-SPECT setup for pinhole pitch of 0.96 mm, pinhole radius of 0.096mm and object to detector gap of 25mm

| Type of L-SPECT | Center Shifting(mm) | FWHM (mm) |
|---|---|---|
| MPRD L-SPECT | NA | 9.48 |
| MPRD L-SPECT with Center Shifting | 5mm | 5.05 |
| MPRD L-SPECT with Center Shifting | 10mm | 5.01 |

From Fig. 7(a) and Fig. 7(b), it can be deduced that MPRD L-SPECT with central shift removes the central artefacts. From visual inspection, it is evident that reducing the module width improves reconstruction quality.

*C. Spatial resolution analysis*

The reconstruction quality was analyzed numerically by using the spatial resolution. It is calculated in terms of the FWHM of the PSF. A spherical point source of 0.2mm in radius is used for the PSF calculation. The PSFs of the LSPECT system with a planar detector, MPRD detector with and without scanning center shift are shown in Fig. 8. Table 1 compares the FWHM of the MPRD L-SPECT with different values of shift to scanning center. This suggests that, for a fixed pinhole pitch and radius, the higher the scanning center shift, the better will be the FWHM. The MTF gives the frequency response of the imaging system.

*D. Comparison of modulation transfer functions*

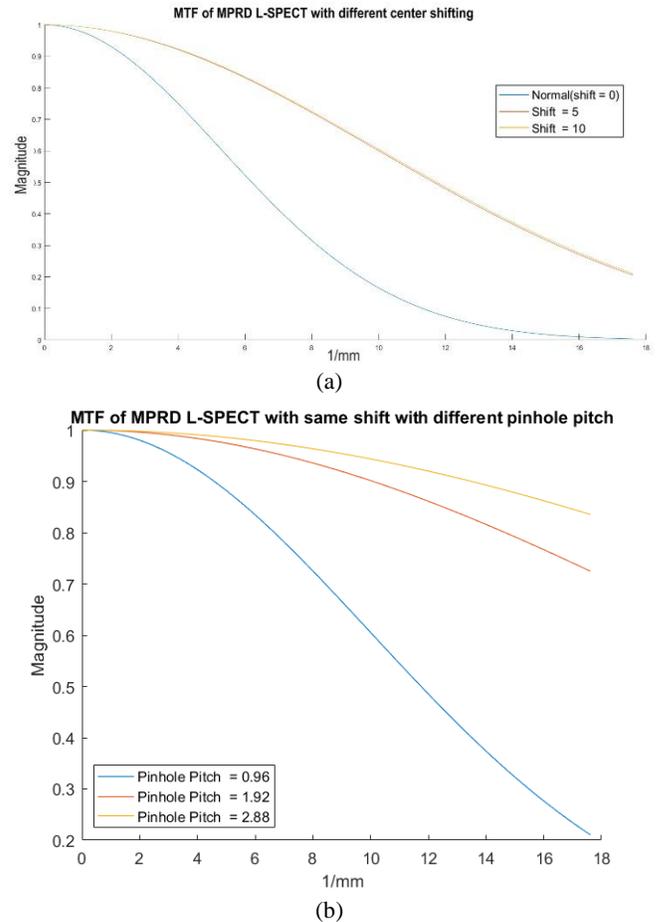

(a)

(b)

Fig. 9 MTF of a) MPRD L-SPECT with different center shifting b) MPRD L-SPECT with same center shifting and different pinhole pitch.

Fig. 9(a) shows the frequency response of the modular partial ring detectors with fixed pinhole pitch and different scanning center shift, and it is clear that the high frequency response will be improved by increasing the shift. Fig 9(b) suggests that even with a center shift, the high frequency response can be further improved by reducing the pinhole pitch.

## IV. DISCUSSION AND CONCLUSION

The L-SPECT system was introduced to improve the drawbacks of conventional SPECT systems, such as low spatial resolution and low sensitivity. In the L-SPECT system, the conventional collimators are replaced with micro range multi-pinhole arrays. This multi-pinhole array improves sensitivity by capturing more information from the radiation sources. The optimal design of the pinhole collimator structure will further improve quality and reduce the number of samples for reconstruction. The modular partial ring detector developed by tiling pinholes and detectors is introduced to improve the FOV and quality of the reconstruction. However, some central artefacts can be found in the case of MPRD L-SPECT after reconstruction. This work focused on reducing the central artefacts of MPRD L-SPECT by shifting the scanning center of the detector. A modified Monte Carlo simulation method is used to generate simulated projection data. The scanned object is reconstructed from projections by raytracing through a cube of interest. The reconstruction quality was analyzed both visually and numerically. From the visual comparison of MPRD L-SPECT, with and without scanning center shift, it is evident that the central artefact was removed by shifting the scanning center. The MTF and FWHM of the point source response are used for numerical comparisons. The analysis of different configurations show that the pinhole pitch and the module width have a significant effect on the reconstruction quality. For the scope of future work, the consideration of probabilistic methods can be used to identify pinhole locations during reconstruction, which can further improve the reconstruction quality of the proposed system.